\documentclass[10pt,letterpaper,english]{article}
\usepackage[T1]{fontenc}
\usepackage[latin1]{inputenc}
\usepackage{amsmath}
\usepackage{graphicx}
\usepackage{amssymb}
\usepackage{array}
\makeatletter

\makeatletter
\usepackage{amscd}

\usepackage{amsfonts}

\usepackage{babel}\makeatother

\makeatother

\usepackage{babel}

\begin{document}
\newtheorem{proposition}{Proposition}[section] \newtheorem{definition}{Definition}[section]
\newtheorem{corollary}{Corollary}[section] \newtheorem{lemma}{Lemma}[section]
\newtheorem{theorem}{Theorem}[section] \newtheorem{example}{Example}[section]

\title{\textbf{Three little arbitrage theorems.}}

\author{Mauricio Contreras G\thanks{Universidad Metropolitana de Ciencias de la Educación UMCE, Chile, email: mauricio.contreras@umce.cl}, \ \ Roberto Ortiz H.\thanks{Universidad Diego Portales, Facultad de Ingeniería, Chile, email: roberto.ortiz@udp.cl} \thanks{Universidad Católica de la Santísima Concepción, FACEA, Chile, email: roberto.ortiz@ucsc.cl} }

\maketitle
\noindent
We prove three theorems about the exact solutions of a generalized or interacting Black--Scholes equation that explicitly includes arbitrage bubbles.
These arbitrage bubbles can be characterized by an arbitrage number $A_N(T)$.
The first theorem states that if $A_N(T) = 0$, then the solution at the maturity of the interacting equation is identical to the solution of the free Black--Scholes equation with the same initial interest rate $r$.
The second theorem states that if $A_N(T) \ne 0$, the solution can be expressed in terms of all higher derivatives of solutions to the free Black--Scholes equation with the initial interest rate $r$.
The third theorem states that whatever the arbitrage number is, the solution is a solution to the free Black--Scholes equation with a variable interest rate $r(\tau) = r + (1/\tau) A_N(\tau)$.
Also, we show, by using the Feynman--Kac theorem, that for the special case of a Call contract, the exact solution for a Call with strike price $K$ is equivalent to the usual Call solution to the Black--Scholes equation with strike price $\tilde{K} = K e^{-A_N(T)}$.\\ \\ \\

\newpage
\section{Introduction}
\noindent

Since its introduction in 1973, the initial Black--Scholes model \cite{bs}, \cite{merton} has undergone many changes over time, which has given rise to many different financial models, such as: stochastic volatility models \cite{heston}--\cite{gatheral} and the associated concept of volatility smile \cite{wilmott}--\cite{dupire}; stochastic rate models \cite{wilmott}, \cite{bjork}--\cite{ho} which account for the spot interest rate dynamic in the determination of the option pricing, as well as, for example, the incorporation of jumps, something which gives rise to integro-differential equations for the price of the option \cite{cont}.
All these generalizations are related to relaxing some assumptions of the initial Black--Scholes model.
One of the most important of these initial axioms is the non-arbitrage hypothesis, which is associated with an equilibrium market dynamics, for which the portfolio return satisfies the standard non-arbitrage condition 
\begin{equation} 
d P(t) = r P(t) dt ,
\end{equation}
where $P(t)$ is the value of the portfolio at time $t$, and $r$ is the constant risk-free interest rate.\\
When deviations from this equilibrium state are considered (which can be for several external reasons, such as: market imperfections such as transaction costs, asymmetric information issues, short-term volatility, extreme discontinuities, among many others), then the classical non-arbitrage assumption is violated.
Thus, other different types of models try to relax the no-arbitrage hypothesis to incorporate such nonequilibrium behaviour.
In fact, economists have realised that futures contracts are not always traded at a price predicted by the simple no-arbitrage relation in a real market.
Substantial empirical evidence has supported this point many times and in different settings \cite{cornell}--\cite{adrian}.\\ 
Most of the attempts to take into account arbitrage in option pricing assume that the return from the B--S portfolio $P$ is not equal to the constant risk-free interest rate $r$, changing the no-arbitrage principle to an equation of the form
$$
d P(t)=\left(r+x(t)\right) P(t) d t ,
$$
where $x(t)$ is a random arbitrage return.
For example, Ilinski \cite{ilinski1, ilinski2} and Ilinski and Stepanenko \cite{ilinski3} assume that $x(t)$ follows an Ornstein--Uhlenbeck process.\\ 

In \cite{contreras1}, inspired by the ideas of \cite{ilinski1}, a generalization of the Black--Scholes (B--S) model that incorporates market imperfections through the presence of arbitrage bubbles was proposed.
In this case, the portfolio return follows a stochastic dynamics of the form
\begin{equation} \label{dP}
d P(t)=r P(t) dt + f(S,t) P(t) d W ,
\end{equation}
(where the amplitude of the Wiener process $f(S,t)$ is called an arbitrage bubble) and which is similar to the equation that determines the dynamics of the asset price $S$ 
\begin{equation} \label{dS}
d S(t)= \alpha S(t) dt + \sigma S(t) d W .
\end{equation}
where $\alpha$ and $\sigma$ are the constant mean and volatility of the underlying asset price, respectively.\\
In \cite{contreras1}, after analysing empirical financial data, it was shown that the mispricing between the empirical and the Black--Scholes prices can be described by a Heaviside type function in time.
This implies that the arbitrage bubble $f$ has the same time dependence.
Thus, arbitrage bubbles can be characterised by a finite time-span and a constant height that measures the Black-Scholes model's price deviation.
Note that the arbitrage bubble can be determined approximately from the empirical financial data, as in  \cite{contreras2}, by using semi-classical methods \cite{contreras3}.
Also, a stochastic generalization of this bubble model has been proposed in \cite{contreras6}.\\ \\
Equations (\ref{dP}) and (\ref{dS}) imply \cite{contreras1}  that the Black--Scholes equation in the presence of an arbitrage bubble is given by 
\begin{equation} \label{BSE1}
\frac{\partial \pi}{\partial t}+\frac{1}{2} \sigma^{2} S^{2} \frac{\partial^{2} \pi}{\partial S^{2}}+r \frac{\left(\sigma-\frac{\alpha f(S,t)}{r}\right)}{(\sigma-f(S,t))}\left(S \frac{\partial \pi}{\partial S}-\pi\right) = 0.
\end{equation}
The above equation can written  as
\begin{equation} \label{BSEinteraction}
\frac{\partial \pi}{\partial t}+\frac{1}{2} \sigma^{2} S^{2} \frac{\partial^{2} \pi}{\partial S^{2}}+r\left(S \frac{\partial \pi}{\partial S}-\pi\right) + v(S,t) \left(S \frac{\partial \pi}{\partial S}-\pi\right)=0,
\end{equation}
with
\begin{equation} \label{potentialSt}
v(S,t) = \frac{(r-\alpha) f(S,t)}{\sigma-f(S,t)},
\end{equation}
which can be interpreted as the potential of an external time-dependent force generated by the arbitrage bubble $f(S,t)$.
Note that if $f = 0$, the potential $v =0$ and so (\ref{BSEinteraction}) reduces to the usual free Black--Scholes model.
Then, under market imperfections, the free Black--Scholes model becomes an interacting one and (\ref{BSEinteraction}) represents a non-martingale generalization of the Black--Scholes equation.\\  

Thus, a generic way to represent a non-risk-free portfolio is given by Equation (\ref{dP}), where $f(S,t)$ encapsulates all of the information about the market equilibrium's deviations no matter what their causes are. Then, in principle, any non-equilibrium option behavior can be modeled endogenously in the framework of Equation (\ref{BSEinteraction}).\\

\section{A quantum mechanical interpretation}
Consider again the Black--Scholes equation (\ref{BSEinteraction}) in the presence of an arbitrage bubble $f(S,t)$.
This equation must be integrated with the final condition
\begin{equation}
\pi(S,T) = \Phi(S).
\end{equation}
The function $\Phi$ is called the contract function and defines the type of option.
Note that Equation (\ref{BSEinteraction}) must be integrated backward in time from the future time $t=T$ to the present time $t=0$.
One can change the direction of time by using the change of variables given by
\begin{equation}
\tau = T - t ,
\end{equation}
which implies that
\begin{equation}
\frac{\partial }{\partial \tau}  = - \frac{\partial}{\partial t} ,
\end{equation}
so (\ref{BSEinteraction}) can be written as forward $\tau$ time Euclidean Schr\"{o}dinger like equation \cite{Merzbacher}, \cite{Sakurai} 
\begin{equation} \label{BSEinteractionv3}
\frac{\partial \pi}{\partial \tau} = \frac{1}{2} \sigma^{2} S^{2} \frac{\partial^{2} \pi}{\partial S^{2}}+r\left(S \frac{\partial \pi}{\partial S}-\pi\right) + v(S, \tau) \left(S \frac{\partial \pi}{\partial S}-\pi\right) ,
\end{equation}
with Hamiltonian operator
\begin{equation} 
\check{H} = \frac{1}{2} \sigma^{2} \check{T} + r \check{P} + v(S, \tau) \check{P},
\end{equation}
where 
\begin{equation} 
\check{T} = S^{2} \frac{\partial^{2} }{\partial S^{2}},
\end{equation}
and
\begin{equation} \label{P_operator}
\check{P} = \left(S \frac{\partial }{\partial S} - \check{I} \right).
\end{equation}
When the amplitude of the bubble is zero, the potential function $v(S, \tau)$ is also zero, and the Hamiltonian reduces to
\begin{equation} \label{H0}
	\check{H}_0 = \frac{1}{2} \sigma^{2} \check{T} + r \check{P}, 
\end{equation}
which gives the evolution of the usual Black--Scholes model.\\

To explore some consequences of Equation (\ref{BSEinteractionv3}), we consider the simplest case of a pure time-dependent arbitrage bubble $f = f(\tau)$, so the potential is $v(\tau) = \frac{(r-\alpha) f(\tau)}{\sigma-f(\tau)}$ and (\ref{BSEinteractionv3}) becomes
\begin{equation} \label{BSEinteractionv5}
\frac{\partial \pi}{\partial \tau} =\check{H}_0 \pi + v(\tau) \check{P} \pi .
\end{equation}
Due to that $\check{P}$ commutes with $ \check{T}$ and $\check{H}_0$, Equation (\ref{BSEinteractionv5}) can be integrated to give
\begin{equation}
\pi(S, \tau) =  e^{ \check{H}_0 \tau + \big( \int_{0}^{\tau} v(\tau^\prime) d\tau^\prime \big) \check{P} } \ \Phi(S),
\end{equation}
where $\pi(S,0) = \Phi(S)$ is the contract function.\\ \\
Now, we define the arbitrage number $A_N(\tau)$ at time $\tau$, associated to the arbitrage  bubble $f(\tau)$, by the integral 
\begin{equation}
A_N(\tau) = \int_{0}^{\tau} v(\tau^\prime) d\tau^\prime = \int_{0}^{\tau} \frac{(r-\alpha) f(\tau^\prime)}{\sigma-f(\tau^\prime)} d\tau^\prime , 
\end{equation}
which represents the accumulated potential between $0$ and $\tau$, so
\begin{equation} \label{pi(S,tau)_interactuante}
\pi(S, \tau) =  e^{ \check{H}_0 \tau  + A_N (\tau) \check{P} }  \ \Phi(S) ,
\end{equation}
that is
\begin{equation}
\pi(S, \tau) =  e^{ A_N (\tau) \check{P} } e^{ \check{H}_0 \tau }  \ \Phi(S) .
\end{equation}
We denote by $C(S,\tau,r)$ the solution to the free Black--Scholes equation for a contract $\Phi(S)$, which evolves with the Hamiltonian (\ref{H0}) characterized by the constant interest rate $r$, so we can write generically
\begin{equation}
C(S, \tau, r) = e^{ \check{H}_0 \tau }  \ \Phi(S) .
\end{equation}
Thus, the solution to the interacting equation is
\begin{equation}
\pi(S, \tau) =  e^{ A_N (\tau) \check{P} } C(S, \tau, r) .
\end{equation}
By using the expansion given in \cite{contreras5} for $e^{ A_N (\tau) \check{P} } $, the last equation can be written as
\begin{equation} \label{perturbative_expansion}
\pi(S, \tau) = \sum_{n=0}^{\infty} e^{-A_N (\tau)} Q_{n}\left(A_N (\tau)\right) S^{n} \ \frac{\partial^n \  }{\partial S^n} C(S, \tau, r) ,
\end{equation}
where 
\begin{eqnarray}
	Q_{0}(x)=&1 &\\
	Q_{1}(x)=&e^{x}-1& \\
	Q_{j}(x)=&\sum_{m=j}^{\infty} \alpha_{m, j} \frac{x^{m}}{m !} & \quad j=2,3,4, \cdots ,
\end{eqnarray}
and the coefficients $\alpha_{m, j}$ are given by the recurrence relation
\begin{equation}
\begin{array}{l}
\alpha_{n, 1}=1,  \ \ \ \ \ \ \alpha_{n, n}=1 \ \ \ \ \ \  \text{and} \\
\alpha_{n m}=m \alpha_{n-1, m}+\alpha_{n-1, m-1} .
\end{array}
\end{equation}
Equation (\ref{perturbative_expansion}) permits writing the solution of the interacting Black-Scholes equation in terms of all the derivatives of the free Black--Scholes solution, that is, in terms of all its Greeks.

\section{The arbitrage theorems}

The arbitrage theorems can be stated as follows. \\  

{\bf First arbitrage theorem}: Let $f(\tau)$ be an arbitrage bubble that acts between $ 0 < \tau < T $.
If the arbitrage number of $f$ at time $\tau = T$ is zero, that is $A_N (T) = 0$, then
\begin{equation} 
\pi(S, T) =  C(S, T, r).
\end{equation}
This means that the solution of the interacting Black--Scholes equation is just the solution to the free Black--Scholes equation at $\tau = T$.\\ \\ 
Proof:  Consider the option price at time $\tau = T$ (or $t=0$), that is,
\begin{equation} \label{perturbative_expansion_at_T}
\pi(S, T) = \sum_{n=0}^{\infty} e^{-A_N (T)} Q_{n}\left(A_N (T)\right) S^{n} \ \frac{\partial^n \  }{\partial S^n} C(S, T, r) .
\end{equation}
Because $A_N(T)=0$, 
\begin{equation} 
\pi(S, T) = \sum_{n=0}^{\infty} Q_{n}\left(0\right) S^{n} \ \frac{\partial^n \  }{\partial S^n} C(S, T, r) ,
\end{equation}
and by noting that the functions $Q_n(x)$ satisfy 
$Q_n(0)=0$ for $n = 1, 2, 3, ...$  and $Q_0(x)=1$, we have
\begin{equation} 
\pi(S, T) = C(S, T, r).
\end{equation}
\noindent \\
{\bf Second arbitrage theorem}: Let $f(\tau)$ be an arbitrage bubble that acts between $ 0 < \tau < T $.
If the arbitrage number of $f$ at time $\tau = T$ is nonzero, that is, $A_N (T) \ne 0$, then the solution of the interacting Black--Scholes equation depends on all Greeks and the bubble's behavior for $0 < \tau < T$, and 
\begin{equation} 
\pi(S, T) = \sum_{n=0}^{\infty} e^{-A_N (T)} Q_{n}\left(A_N (T)\right) S^{n} \ \frac{\partial^n \  }{\partial S^n} C(S, T, r) .
\end{equation}
Proof: Evaluate $\pi(S, \tau$) in  (\ref{perturbative_expansion}) at $ \tau = T$.
\noindent \\\\

{\bf Third arbitrage theorem}:  Let $f(\tau)$ be an arbitrage bubble that acts between $ 0 < \tau < T $ and let $A_N (\tau)$ be the arbitrage number of $f$ at time $\tau$. Then the solution $\pi(S, \tau)$ of the interacting Black--Scholes equation is just the solution to a free Black--Scholes equation with the variable interest rate
$$r(\tau)=r + \frac{1}{\tau} A_N(\tau) ,$$
that is,
\begin{equation} 
 \pi(S, \tau) = C(S, \tau, r(\tau)) ,
\end{equation}
and
\begin{equation} \label{identityCexpansion}
C(S, \tau, r + \frac{1}{\tau} A_N(\tau) ) =  \sum_{n=0}^{\infty} e^{-A_N (\tau)} Q_{n}\left(A_N (\tau)\right) S^{n} \ \frac{\partial^n \  }{\partial S^n} C(S, \tau, r) .
\end{equation} \\
Proof: Consider Equation (\ref{pi(S,tau)_interactuante}), which is equivalent to
\begin{equation} \label{pi(S,tau)_interactuante_2}
\pi(S, \tau) =  e^{ \check{H}_0 \tau  + \frac{1}{\tau} A_N (\tau)  \check{P} \tau }  \ \Phi(S) =  e^{ \left( \frac{1}{2} \sigma^{2} \check{T} + r \check{P} \right) \tau  + \frac{1}{\tau} A_N (\tau) \check{P} \tau }  \ \Phi(S) ,
\end{equation}
that is
\begin{equation} \label{pi(S,tau)_interactuante_3}
\pi(S, \tau) =  e^{ \left( \frac{1}{2} \sigma^{2} \check{T} + (r+ \frac{1}{\tau} A_N (\tau) ) \check{P} \right) \tau }  \ \Phi(S) =e^{ \left( \frac{1}{2} \sigma^{2} \check{T} + r(\tau) \check{P} \right) \tau }  \ \Phi(S) =  C(S, \tau, r(\tau)) ,
\end{equation}
where $r(\tau) = r+ \frac{1}{\tau} A_N (\tau)$.
By comparing (\ref{pi(S,tau)_interactuante_3}) with (\ref{perturbative_expansion}) we obtain (\ref{identityCexpansion}). \\

\section{A stochastic interpretation}
A second way to prove the first arbitrage theorem is to consider the Feynman--Kac theorem \cite{bjork}.
In such case, the solution $\pi(S, t)$ of the interacting Black--Scholes equation (\ref{BSEinteraction}) can be represented as
\begin{equation}
\pi(S, t) = \ e^{ -r(T-t) - \int_{t}^{T} v(t^{\prime}) d t^{\prime} } \ \mathbb{E} \Big[  \Phi(S_T) \Big] ,
\end{equation}
where
\begin{equation}
S_T = S \ e^{ ( r - \frac{1}{2} \sigma^2 ) (T-t) + \int_{t}^{T} v(t^\prime) dt^\prime + \sigma (W_T -W_{t}) } .
\end{equation}
In particular, for $t=0$
\begin{equation} \label{FKT_1}
\pi(S, 0) = \ e^{ -rT - A_N(T) } \ \mathbb{E} \Big[  \Phi(S_T) \Big] ,
\end{equation}
with
\begin{equation} \label{FKT_2}
S_T = S \ e^{ ( r - \frac{1}{2} \sigma^2 ) T + A_N(T) + \sigma W_T } .
\end{equation}
For a bubble with zero arbitrage number one gets
\begin{equation}
\pi(S, 0) = \ e^{ -rT } \ \mathbb{E} \Big[  \Phi(S_T) \Big] ,
\end{equation}
\begin{equation}
S_T = S \ e^{ ( r - \frac{1}{2} \sigma^2 ) T + \sigma W_T } ,
\end{equation}
which is precisely the solution of the free Black-Scholes model.
For a particular case of a Call, $\Phi(S) = \max \{0 , S - K  \} $ so (\ref{FKT_1}) and  (\ref{FKT_2}) gives
\begin{equation} \label{FKT_1b}
\pi(S, 0) = \ e^{ -rT - A_N(T) } \ \mathbb{E} \Big[  \max \{0 ,  S \ e^{ ( r - \frac{1}{2} \sigma^2 ) T + A_N(T) + \sigma W_T } - K  \} \Big] ,
\end{equation}
or
\begin{equation} \label{FKT_1c}
\pi(S, 0) = \ e^{ -rT } \ \mathbb{E} \Big[  \max \{0 ,  S \ e^{ ( r - \frac{1}{2} \sigma^2 ) T + \sigma W_T } - K e^{- A_N(T) } \} \Big].
\end{equation}
Thus, for a Call contract, the solution of the interacting Black--Scholes with strike price $K$ is equivalent to the solution of the free Black--Scholes equation with strike price $\tilde{K}$ given by
\begin{equation} 
\tilde{K} = K e^{- A_N(T) } ,
\end{equation}
which depends of the bubble's past dynamics.
Of course, for a $A_N(T)=0$, $ \tilde{K} = K$. Note that for $A_N(T) \rightarrow + \infty$ or $A_N(T) \rightarrow - \infty$ (that is, in the resonant case reported in \cite{contreras4}), the option price goes to that a Call with zero strike price ($ \tilde{K} = 0 $) or the option price is zero for all $S$ ($\tilde{K} = \infty$) respectively. \\

\section{Some applications}
To illustrate the consequences of the two above theorems, we consider the simple single square-time bubble form \cite{contreras1}
\begin{equation}
f(\tau)=\left\{\begin{array}{ll}
0 & 0 \le \tau \le \tau_{1} \\
f_0 & \tau_{1} < \tau < \tau_{2} \\
0 & \tau_{2} \le  \tau \le T
\end{array}\right.
,
\end{equation}
where $\tau_1 = T - T_2$ and $\tau_2 = T - T_1$.
The potential function (\ref{potentialSt}) for this bubble is 
\begin{equation} \label{potentialfunction2}
v(\tau) = \left\{\begin{array}{ll}
0 & 0 \le \tau \le \tau_{1} \\
v_0 = \frac{(r-\alpha) f_0}{\sigma-f_0}, & \tau_{1} < \tau < \tau_{2} \\
0 & \tau_{2} \le \tau \le T
\end{array}\right.
,
\end{equation}
and the arbitrage number for this square bubble is
\begin{equation}
A_N(T) = \int_{0}^{T} v(\tau^\prime) d\tau^\prime = \int_{\tau_1}^{\tau_2} \frac{(r-\alpha) f_0}{\sigma-f_0} d\tau^\prime = v_0 (\tau_2 - \tau_1).
\end{equation} \\
In this case the interacting solution is given by
\begin{equation} 
\pi(S, \tau) = \left\{\begin{array}{ll}
C\big(S, \tau, r \big) & 0 \le \tau \le \tau_{1} \\
C\big(S, \tau, r + \frac{ v_0 (\tau - \tau_1)}{\tau} \big) & \tau_{1} < \tau < \tau_{2} \\
C\big(S, \tau, r + \frac{ v_0 (\tau_2 - \tau_1)}{\tau} \big)  & \tau_{2} \le \tau \le T
\end{array}\right.
.
\end{equation}

When $r < \alpha$ and $f_0 > \sigma$, we speak of positive bubble because the arbitrage number is positive.
For $f_0 < \sigma$, the arbitrage number becomes negative, so we speak of a negative bubble.
For the case $r > \alpha$ we have a positive bubble for $f_0 < \sigma$ and negative bubble for $f_0 > \sigma$.
Note that a single positive or negative square bubble has an arbitrage number $A_N(T) \ne 0$.\\ \\ 
Consider now the arbitrage process generated successively by a pair of positive and negative bubbles
\begin{equation}
f(\tau)=\left\{\begin{array}{ll}
0 & 0 \le \tau \le \tau_{1} \\
f_0 & \tau_{1} < \tau < \tau_{2} \\
f_1 & \tau_{2} < \tau < \tau_{3} \\
0 & \tau_{3} \le  \tau \le T .
\end{array}\right.
\end{equation} 
If the total arbitrage number of the positive and negative bubbles is zero, then we speak of a bubble--antibubble pair. \\\\

Figure 1 shows a bubble--antibubble pair $f$ for a Call with $ K= 5,\ r = 0.1, \ \sigma = 0.3,\ T = 1$ and $ \alpha = 0.2 $, $ \tau_1 =0.4, \ \tau_2 =0.5, \ \tau_3 =0.6 $.
Here, $ f_0 = 0.285 $ for the negative bubble and $ f_1 = 0.3167 $ for the positive bubble.
\begin{figure}[h!]
	\centering
	\includegraphics[scale=0.7]{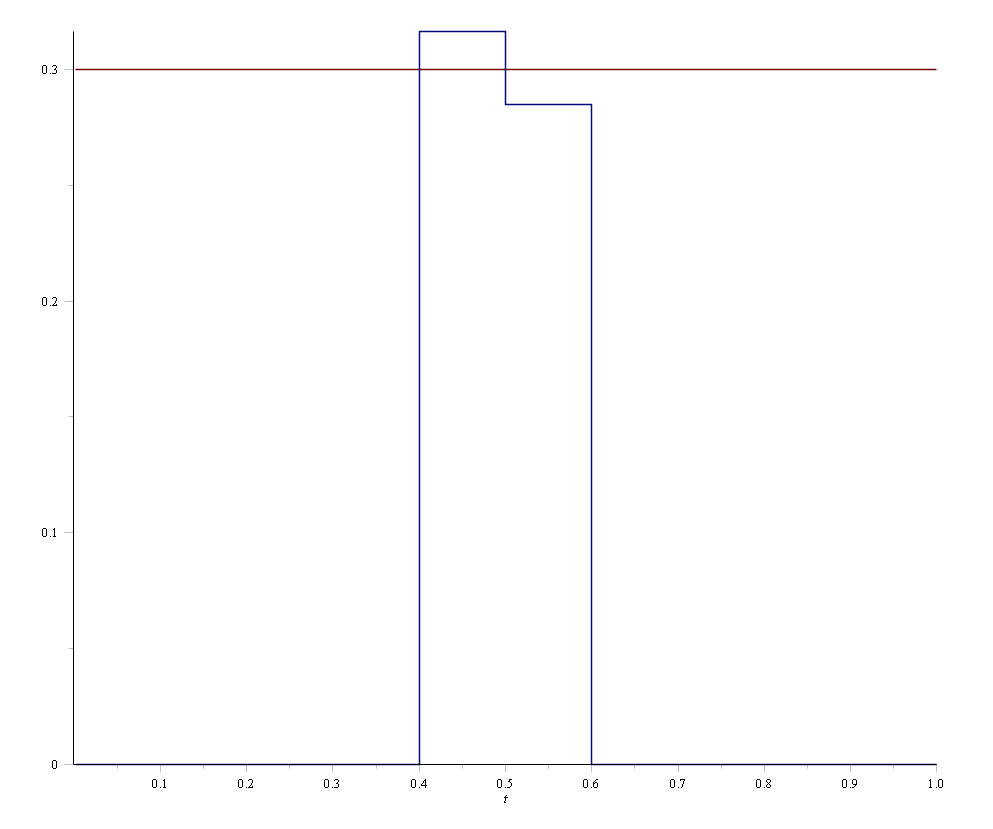}
	\includegraphics[scale=0.7]{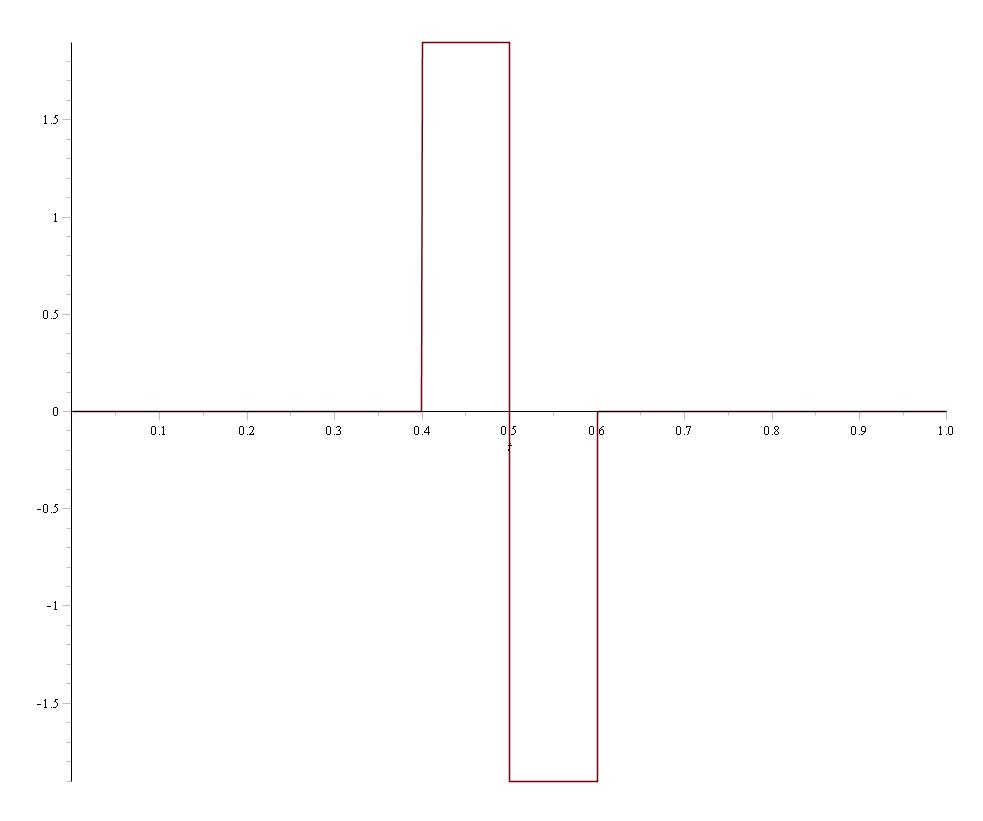}
	\includegraphics[scale=0.7]{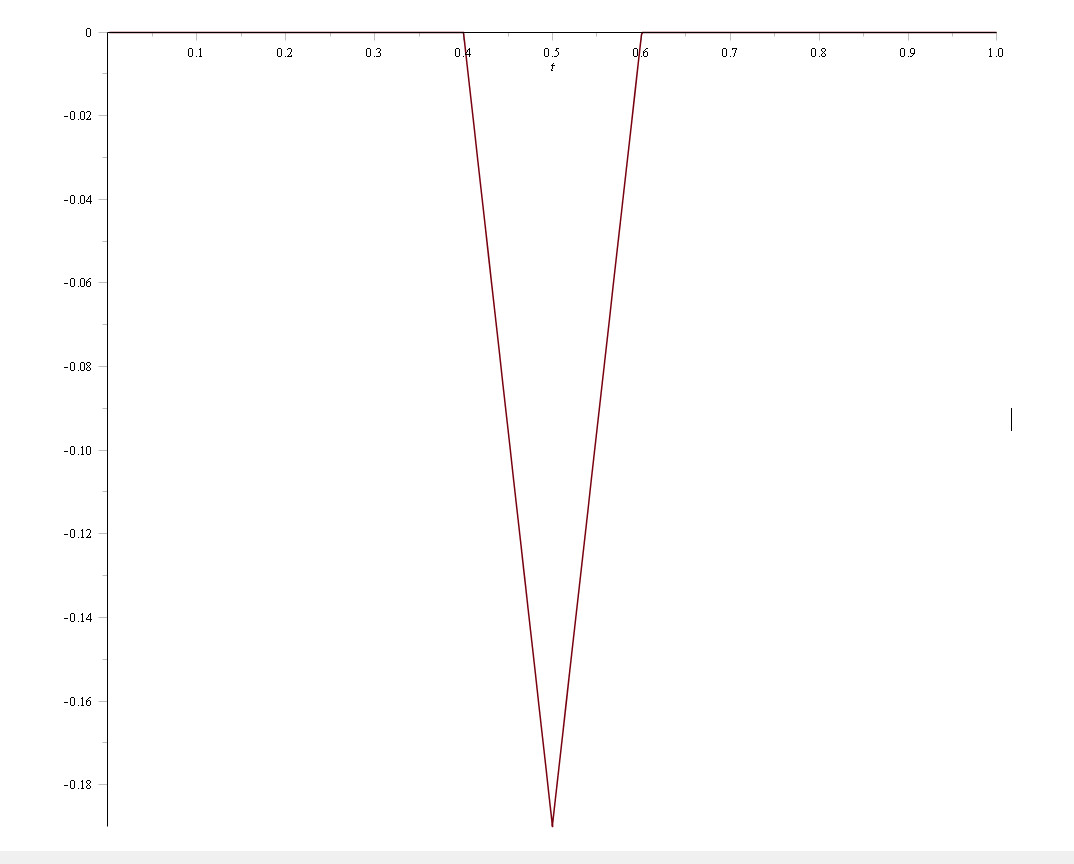}
	\caption{Upper left: blue curve, bubble--antibubble pair as a function of time $t=T-\tau$.
Red curve, the volatility $\sigma$.
Note that the positive and negative bubbles are equidistant with respect to $\sigma$.
Upper right: potential $v$ in terms of $t$.
Lower figure: arbitrage number in terms of $t$.
Note that after the bubble--antibubble pair has finished acting, the arbitrage number is zero.}
\end{figure} \\ \\
Figure 2 shows the effect of a bubble--antibubble pair on the option price dynamic.
The curve at $t=0$ is the usual Black--Scholes solution with interest rate $r$.\\
\begin{figure}[h!]
	\centering
	\includegraphics[scale=0.7]{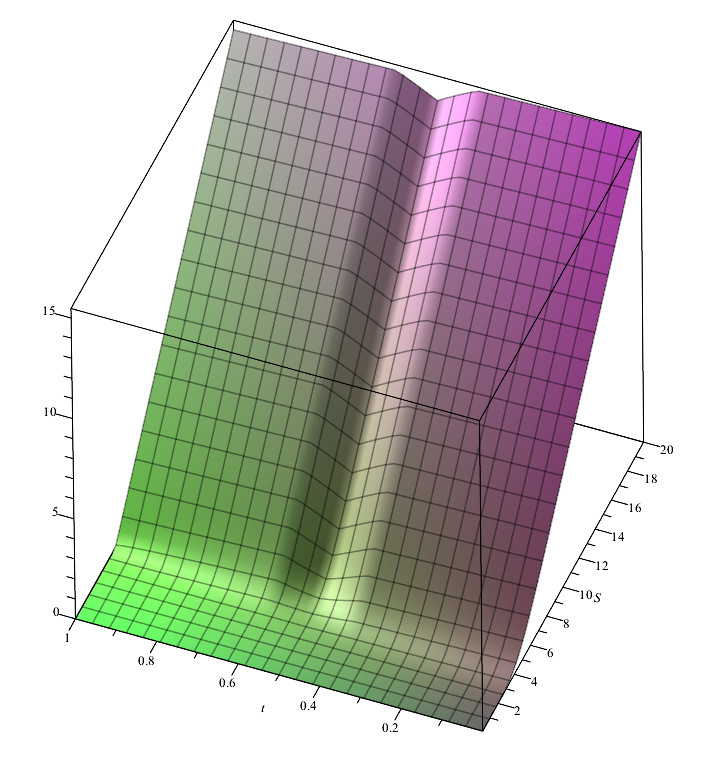}
	\caption{Left: effect of a bubble--antibubble pair on the option price.}
\end{figure} \\

\noindent
\newpage
Now consider the case of three successive square bubbles of the form
\begin{equation}
f(\tau)=\left\{\begin{array}{ll}
0 & 0 \le \tau \le \tau_{1} \\
f_0 & \tau_{1} < \tau < \tau_{2} \\
f_1 & \tau_{2} < \tau < \tau_{3} \\
f_2 & \tau_{3} < \tau < \tau_{4} \\
0 & \tau_{4} \le  \tau \le T. 
\end{array}\right.
\end{equation}  
Figure 3 shows the effect of  two negative bubbles and one positive bubble (left figure) as well as that of one positive and two negative bubbles (right figure).
In all cases the total arbitrage number is zero.
Here $ K= 5,\ r = 0.1, \ \sigma = 0.3,\ T = 1$ and $ \alpha = 0.2 $, $ \tau_1 =0.2, \ \tau_2 =0.4, \ \tau_3 =0.5, \ \tau_4 =0.6 $.
For the left figure $ f_0 = 0.24, \ f_1=0.276, \ f_2=0.316 $, whereas for the right figure $ f_0 = 0.33, \ f_1=0.24, \ f_2=0.28 $.
\begin{figure}[h!]
	\centering
	\includegraphics[scale=0.7]{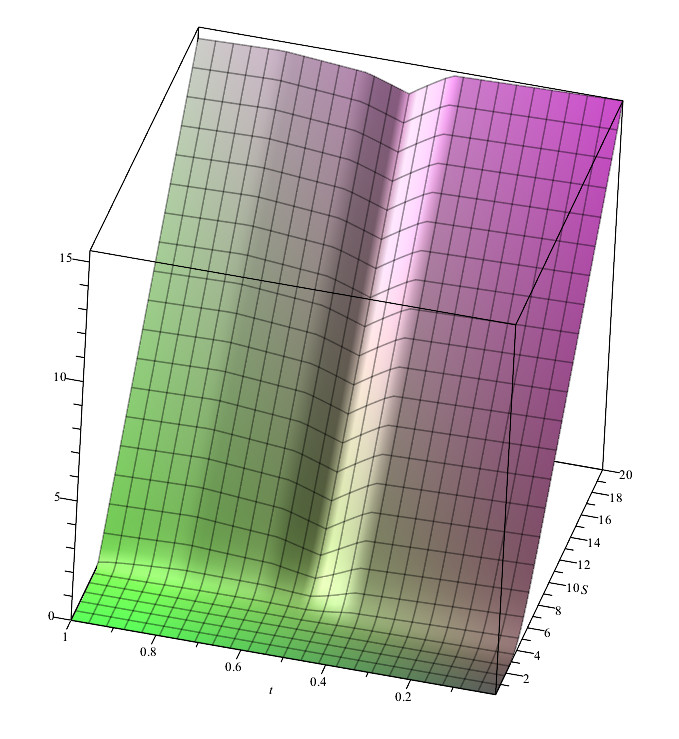}
	\includegraphics[scale=0.7]{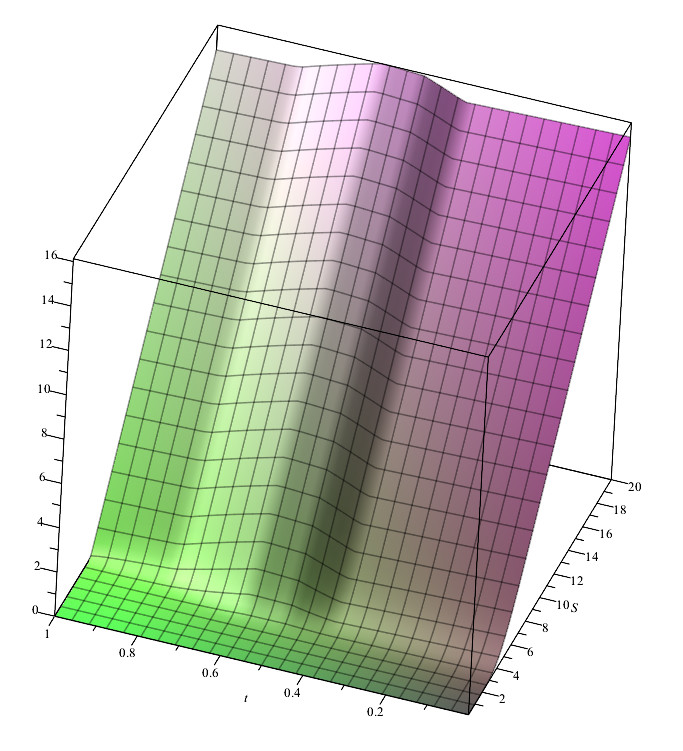}
	\caption{Left: Effect of two negative and one positive bubbles respectively; Right: Effect of one positive and two negative bubbles.
In all cases the total arbitrage number is zero.}
\end{figure} \\\\
Figure 4 shows the effect of one positive, one negative and one positive bubbles respectively (left figure), and one negative, one positive and one negative bubbles respectively (right figure).
In all cases the total arbitrage number is zero.
Here $ K= 5,\ r = 0.1, \ \sigma = 0.3,\ T = 1$ and $ \alpha = 0.2 $, $ \tau_1 =0.2, \ \tau_2 =0.3, \ \tau_3 =0.5, \ \tau_4 =0.6 $.
For the left figure $ f_0 = 0.34, \ f_1=0.273, \ f_2=0.326 $, whereas for the right figure $ f_0 = 0.27, \ f_1=0.33, \ f_2=0.278 $.
\begin{figure}[h!]
	\centering
	\includegraphics[scale=0.7]{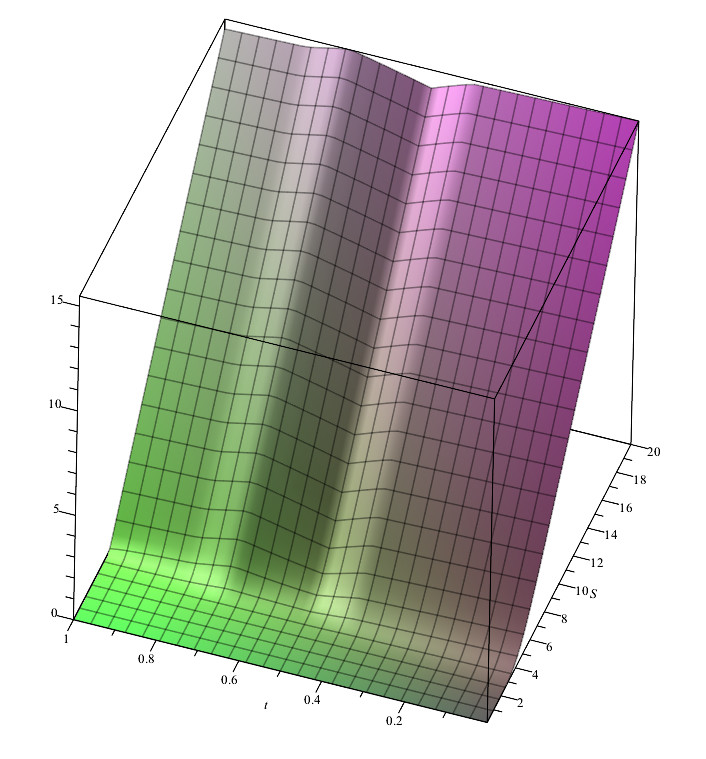}
	\includegraphics[scale=0.7]{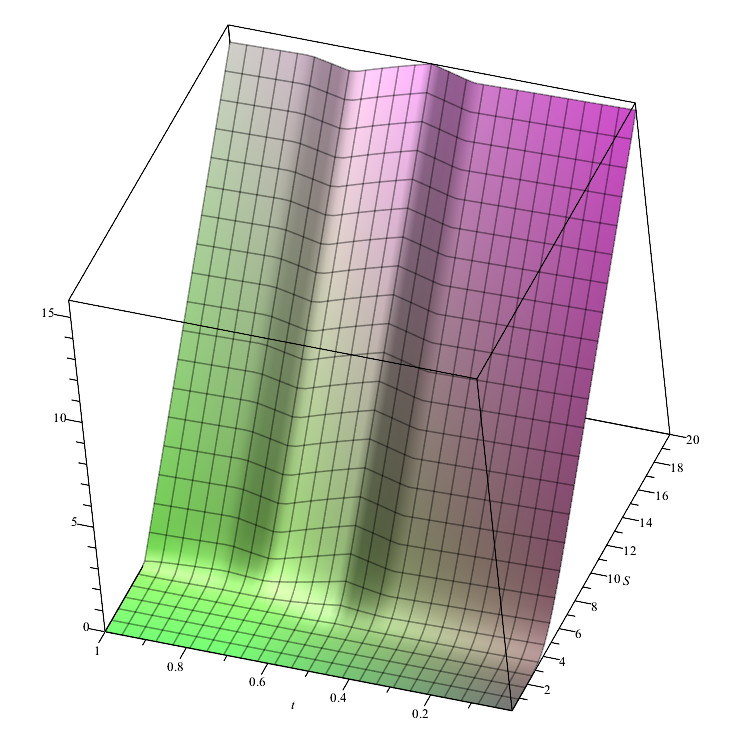}
	\caption{Left: One positive, one negative and one positive bubble. Right: one negative, one positive and one negative bubble.
The total arbitrage number is zero.}
\end{figure} 

\noindent
\newpage
Figure 5 shows the effect of two negative and one positive bubble, with total arbitrage number $ A_N \ne 0 $.\\\\
\begin{figure}[h!]
	\centering
	\includegraphics[scale=0.7]{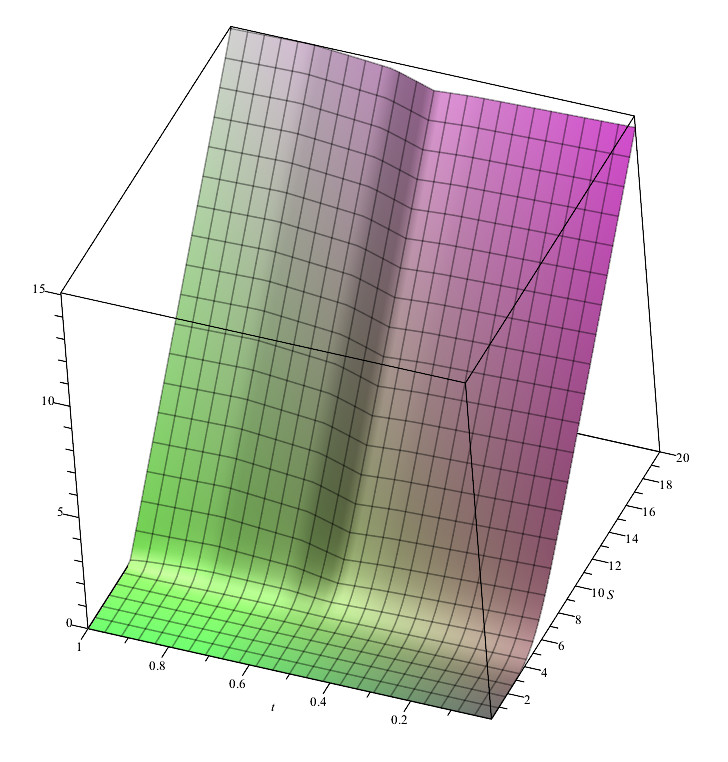}
	\includegraphics[scale=0.7]{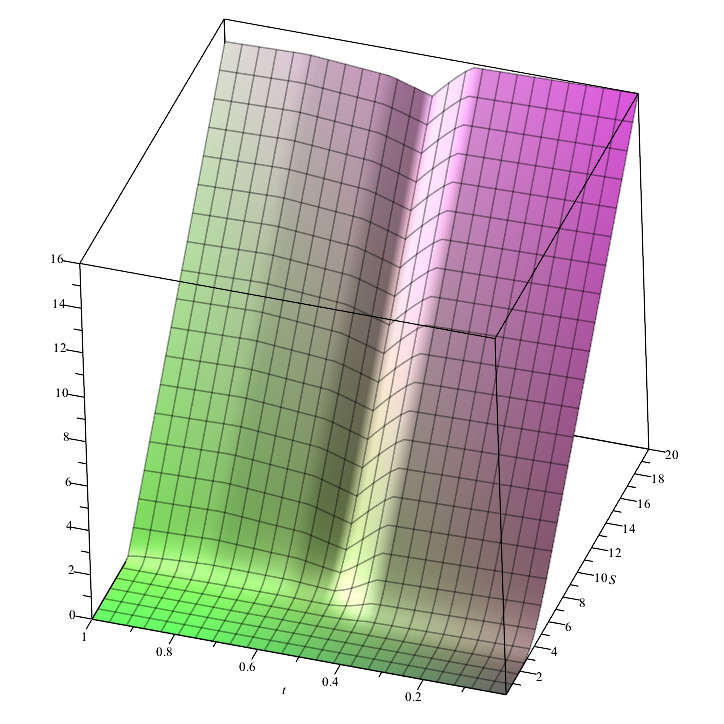}
	\caption{Left: Effect of two negative and one positive bubble, with total arbitrage number $A_N = -0.18$; Right: Same but with $A_N = 0.12$.}
\end{figure}  
\noindent  \\

\section{Conclusions}
Three theorems about the exact dynamical behaviour of the option price when time-dependent arbitrage bubbles are incorporated explicitly in the Black--Scholes equation have been presented.
These arbitrage bubbles (which can act at some instant $t$ between $0$ and the maturity $T$) can be characterized by an arbitrage number $A_N (\tau)$ that corresponds to the accumulated external potential from $0$ to time $\tau$.
If $A_N(T) = 0$, independently of the bubble's dynamic behaviour, the option's values at its maturity are given by the usual Black--Scholes dynamics with interest rate $r$.
In some sense, the option does not remember  the past arbitrage process.
If $A_N(T) \ne 0$, the option price depends on the bubble and on all higher derivatives (or Greeks) of the solution of the free Black--Scholes equation with interest rate $r$, and is equal to a solution of the free Black--Scholes equation with interest rate $ r+ (1/T) A_N(T)$.
In this case, the option remembers the past arbitrage processes.
Also, for a special case of a Call contract with strike price $K$, the solution of the interacting Black--Scholes equation is just the usual free Call solution, but with strike price $\tilde{K} = e^{-A_N(T)}$.\\
Thus arbitrage bubbles with $A_N(T) =0$ are innocuous because the system returns to the initial equilibrium state. Instead, bubbles with 
$A_N(T) \ne 0$ are dangerous from a financial point of view, because these bubbles change the initial equilibrium trajectory of the option price. \\
We hope that these theorems will help in understanding the option's dynamic evolution when arbitrage process are included. A generalization of these theorems and consequences, for the case of a price dependent bubble $f=f(S,t)$, will be addressed in future research.\\

\end{document}